\documentclass{article}

     \usepackage[nonatbib,final]{neurips_2020}

\usepackage[utf8]{inputenc} % allow utf-8 input
\usepackage[T1]{fontenc}    % use 8-bit T1 fonts
\usepackage{hyperref}       % hyperlinks
\usepackage{url}            % simple URL typesetting
\usepackage{booktabs}       % professional-quality tables
\usepackage{amsfonts}       % blackboard math symbols
\usepackage{nicefrac}       % compact symbols for 1/2, etc.
\usepackage{microtype}      % microtypography
\usepackage{graphicx}
\usepackage{authblk}
\usepackage{xcolor}
\newcommand{\xhdr}[1]{\noindent{{\bf #1.}}}

\newcommand{\hide}[1]{} %hide

\title{MolDesigner: Interactive Design of Efficacious \\Drugs with Deep Learning}

\hide{
\author[1]{\textbf{Kexin Huang}}
\author[2]{\textbf{Tianfan Fu}} 
\author[3]{\textbf{Dawood Khan}}
\author[3]{\textbf{Ali Abid}}
\author[3]{\textbf{Ali Abdalla}}
\author[3]{\textbf{Abubakar Abid}}
\author[4]{\textbf{Lucas M. Glass}}
\author[1]{\textbf{Marinka Zitnik}}
\author[4]{\textbf{Cao Xiao}}  
\author[5]{\textbf{Jimeng Sun}} 
\affil[1]{\small{Harvard University}}
\affil[2]{Georgia Institute of Technology}
\affil[3]{Gradio}
\affil[4]{IQVIA}
\affil[5]{University of Illinois at Urbana-Champaign}
\affil[]{{\texttt{kexinhuang@hsph.harvard.edu, tfu42@gatech.edu,\{dawood,ali.abid,ali.abdalla,abubakar\}@gradio.app, lucas.glass@iqvia.com, marinka@hms.harvard.edu, cao.xiao@iqvia.com, jimeng@illinois.edu}}}
}

\author{%
  \bf Kexin Huang$^1$, Tianfan Fu$^2$, Dawood Khan$^3$, Ali Abid$^3$,
  Ali Abdalla$^3$, Abubakar Abid$^3$, Lucas M. Glass$^4$, Marinka Zitnik$^1$, Cao Xiao$^4$, Jimeng Sun$^5$ \\
    $^1$Harvard University, $^2$Georgia Institute of Technology,
    $^3$Gradio, $^4$IQVIA, $^5$University of Illinois at Urbana-Champaign\\
    \texttt{kexinhuang@hsph.harvard.edu}, \texttt{tfu42@gatech.edu}, \texttt{\{dawood,ali.abid,ali.abdalla,abubakar\}@gradio.app}, \texttt{lucas.glass@iqvia.com}, \texttt{marinka@hms.harvard.edu}, \texttt{cao.xiao@iqvia.com}, \texttt{jimeng@illinois.edu}  
}

\begin{document}

\maketitle

\begin{abstract}
The efficacy of a drug depends on its binding affinity to the therapeutic target and pharmacokinetics. Deep learning (DL) has demonstrated remarkable progress in predicting drug efficacy. We develop MolDesigner, a human-in-the-loop web user-interface (UI), to assist drug developers leverage DL predictions to design more effective drugs. A developer can draw a drug molecule in the interface. In the backend, more than 17 state-of-the-art DL models generate predictions on important indices that are crucial for a drug's efficacy. Based on these predictions, drug developers can edit the drug molecule and reiterate until satisfaction. MolDesigner can make predictions in real-time with a latency of less than a second.
\end{abstract}

A drug's efficacy depends on various factors. Among others, two crucial factors are binding affinity that measures the strength of drug-protein target interactions and pharmacokinetics (PK) indices that measure how the body reacts to the drug~\cite{schenone2013target}. PK indices include absorption, distribution, metabolism, excretion, and toxicity (ADMET). However, to obtain these values in a traditional wet lab setting is notoriously expensive and time-consuming~\cite{dickson2004key}. Deep learning (DL) models have recently shown high accuracy in predicting the binding affinity and various ADMET indices~\cite{kearnes2016modeling,ozturk2018deepdta,huang2020deeppurpose}.

\xhdr{Demonstrated Technology} 
Here, we propose a humans-in-the-loop web user interface (UI) called MolDesigner to help drug developers leverage DL to improve drug designs. Given a drug molecule, MolDesigner predicts various important indices from state-of-the-art DL models, powered by our DeepPurpose library~\cite{huang2020deeppurpose}. They provide real-time feedback to help drug developers modify the molecular structure and iterate on this process until the models start to show the drug candidate has reasonable measures and properties. This way, drug developers can leverage DL models to rapidly design a drug that has not only high efficacy to the target protein but also has ideal ADMET properties. MolDesigner is especially useful for lead optimization where drug developer makes chemical modification on the molecular structure to improve the drug's quality.

In MolDesigner, the user (drug designer) first specifies the target protein (e.g., amino acid sequence) of interest. The user can then begin drawing the initial drug candidate or reading the molecular structure from a file using any popular chemical data format. In less than one second latency, MolDesigner returns predictions, representing binding affinity to the target protein along with predictions for 16 crucial PK indices. The user can then review predictions and make further chemical modifications based on the expert knowledge, thus iteratively refining the drug candidate's structure until the candidate molecule is predicted to have the desired properties. This draw-predict cycle can be repeated as needed until the user is satisfied. The user can also specify the DL model to make predictions. We provide five DL models for binding affinity prediction and three models for each of the PK properties. In total, more than 50 DL models power the backend of MolDesigner. 

\xhdr{Implementation}
We use DeepPurpose~\cite{huang2020deeppurpose} to rapidly generate DL models for drug-target interaction (DTI) and drug property predictions. We then use Gradio~\cite{abid2019gradio} to set up the web UI. We use the BindingDB dataset for DTI prediction and collect 16 datasets from various sources for the PK indices. The PK indices include solubility, lipophilicity, Caco-2, human-intestinal-absorption (HIA), P-gp inhibition, bioavailability, blood-brain-barrier (BBB), plasma-protein-binding rate (PPBR), various CYP inhibitors, half-life, clearance, and clinical toxicity. We provide the following DL models for drug-protein binding affinity prediction. For modeling drugs, we use Message Passing Neural Network (MPNN)~\cite{gilmer2017neural}, Convolutional Neural Network (CNN)~\cite{krizhevsky2012imagenet}, Deep Neural Network (DNN) on Morgan fingerprints~\cite{rogers2010extended}. For modeling proteins, we have CNN and DNN, which are trained on proteins' amino acid composition fingerprints. For drug property prediction, MolDesigner supports MPNN, CNN, and DNN on Morgan fingerprints. The Gradio interface also supports taking screenshots and flagging outputs for easy tracking and future reference.

\begin{figure}[t]
    \vspace{10mm}
    \centering
    \includegraphics[width = \textwidth]{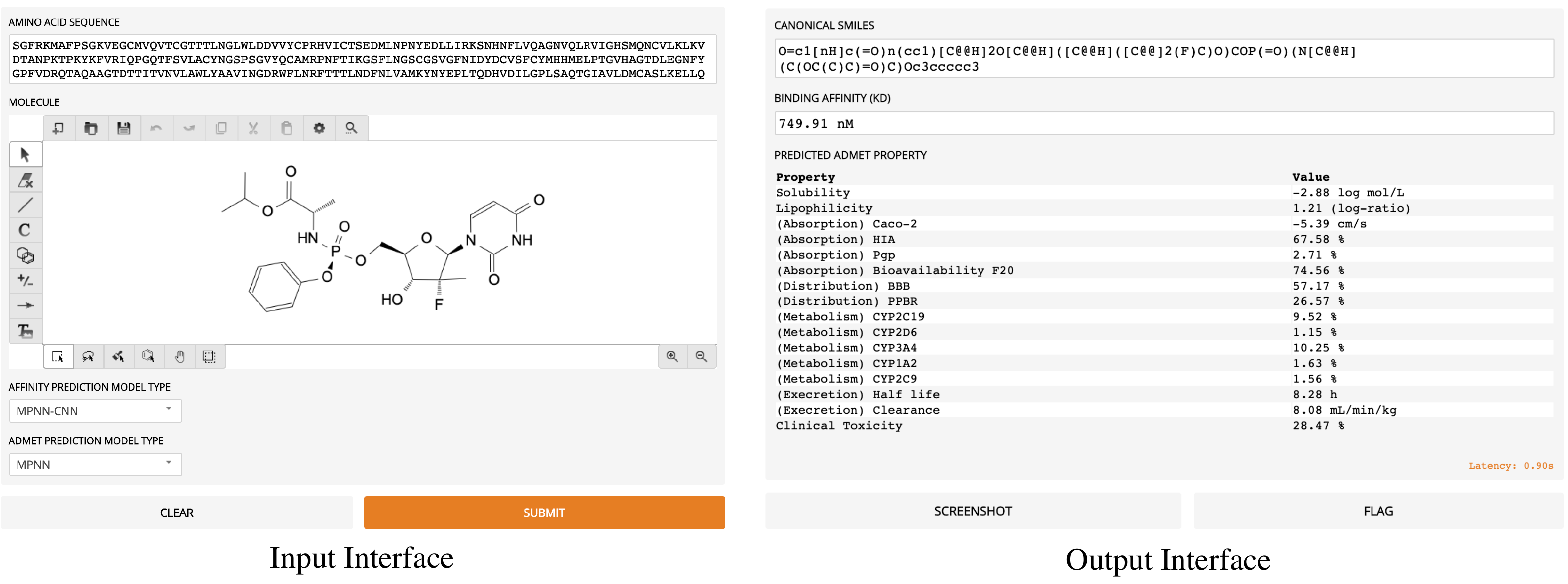}
    \caption{The web user-interface of MolDesigner. }
    \label{fig:my_label}
    \vspace{-7mm}
\end{figure}
\xhdr{Novelty}
MolDesigner makes the following contributions. 1) Existing drug discovery interfaces focus on one task~\cite{dong2018admetlab,alaimo2015dt,chen2016drug}, such as DTI or one of ADMET prediction tasks. In contrast, MolDesigner is the first to integrate 16 ADMET predictions and binding affinity in a single click. 2) Earlier work focuses on a demo of prediction model where a separate output page is generated after user input~\cite{dong2018admetlab,alaimo2015dt,chen2016drug}. In contrast, MolDesigner is designed to keep humans in the loop. MolDesigner focuses on molecular design, enabling rapid predictions, exploration, and interactive adjustments as the user modifies chemical reactions. It provides a unified interface where the user can see both drug molecules and predicted values on the same page. 3) MolDesigner is powered by over 50 state-of-the-art DL models, whereas prevailing tools are only designed to use one model. For ADMET prediction, the current web UI mainly uses classic machine learning methods. 4) MolDesigner is fast. The overall prediction latency is less than one second, allowing rapid molecular design.

\xhdr{Interaction with Virtual Audience} 
We will start with a demo showcasing how our framework aids drug design. Then, depending on the audience's background, we will design two interactive activities. For a computational audience with no knowledge of drug development, we will have a game where users will play the role of chemists developing drugs for one of the COVID-19 targets~\cite{gordon2020sars}. We will start from the drug that already has a high affinity to the target and let the user make chemical modifications. As the user draws drugs, we will demonstrate how predicted drug properties vary with molecular structure. To engage with drug discovery experts in the audience, we will ask them to propose a target of interest and a potential drug molecule.  We will ask them what changes to the molecular structure would they introduce to improve predicted properties, and see if our model can corroborate their hypothesis, followed by discussion. For the audience to engage with the UI, we will use remote control features to draw on the screen. 

\bibliographystyle{plain}
\bibliography{ref}

\end{document}